# Only Aggressive Elephants are Fast Elephants

Jens Dittrich, Jorge-Arnulfo Quiané-Ruiz, Stefan Richter,
Stefan Schuh, Alekh Jindal, Jörg Schad

Information Systems Group
Saarland University
http://infosys.cs.uni-saarland.de

## ABSTRACT

Yellow elephants are slow. A major reason is that they consume their inputs entirely before responding to an elephant rider's orders. Some clever riders have trained their yellow elephants to only consume parts of the inputs before responding. However, the teaching time to make an elephant do that is high. So high that the teaching lessons often do not pay off. We take a different approach. We make elephants aggressive; only this will make them very fast.

We propose HAIL (Hadoop Aggressive Indexing Library), an enhancement of HDFS and Hadoop MapReduce that dramatically improves runtimes of several classes of MapReduce jobs. HAIL changes the upload pipeline of HDFS in order to create different clustered indexes on each data block replica. An interesting feature of HAIL is that we typically create a win-win situation: we improve both data upload to HDFS and the runtime of the actual Hadoop MapReduce job. In terms of data upload, HAIL improves over HDFS by up to 60% with the default replication factor of three. In terms of query execution, we demonstrate that HAIL runs up to 68x faster than Hadoop. In our experiments, we use six clusters including physical and EC2 clusters of up to 100 nodes. A series of scalability experiments also demonstrates the superiority of HAIL.

## 1. INTRODUCTION

MapReduce has become the de facto standard for large scale data processing in many enterprises. It is used for developing novel solutions on massive datasets such as web analytics, relational data analytics, machine learning, data mining, and real-time analytics [16]. In particular, log processing emerges as an important type of data analysis commonly done with MapReduce [3, 24, 13]. Typically, users (as well as developers) want to analyze these web logs in an exploratory way. In fact, Facebook and Twitter use Hadoop MapReduce (the most popular MapReduce open source implementation) to analyze the huge amounts of web logs generated every day by their users [32, 15, 23].

Let us see through the eyes of a representative analyst, say Bob, who wants to analyze a large web log. The web log contains different fields that may serve as filter conditions for Bob like visitDate, adRevenue, sourceIP and so on. Assume Bob is interested in all sourceIPs with a visitDate from 2011. Thus, Bob writes a MapReduce program to filter out exactly those records and discard all others. Bob is using Hadoop. It will scan the entire input dataset from disk to filter out the qualifying records. This takes a while. After inspecting the result set Bob detects a series of strange requests from sourceIP 134.96.223.160. Therefore, he decides to modify his MapReduce job to show all requests from the entire input dataset having that sourceIP. Bob is using Hadoop. This takes a while. Eventually, Bob decides to modify his MapReduce job again to only return log records having a particular adRevenue. Bob is using Hadoop. Yes, it takes a while.

In summary, Bob uses a sequence of different filter conditions, each one triggering a new MapReduce job. He is strolling around. He is not exactly sure what he is looking for. The whole endeavor feels like going shopping without a shopping list: *"let's see what I am going to encounter on the way"*. This kind of use-case illustrates an exploratory usage of Hadoop MapReduce. It is a major use-case of Hadoop MapReduce [3, 13, 26].

This use-case has one major problem: *slow query runtimes.* The time to execute a Hadoop MapReduce job based on a scan may be very high: it is dominated by the I/O for reading all input data [27, 20]. While waiting for his MapReduce job to complete, Bob has enough time to pick a coffee (or two). Every time Bob modifies the MapReduce job, Bob will be able to pick up even more coffee. This increases his caffeine levels to scary heights, kills his productivity, and makes his boss unhappy.

Now, assume the fortunate case that Bob is the type of smart user who thinks a bit about his data before running expensive MapReduce jobs. He has read all the recent VLDB papers (including [8, 12, 22, 25, 17, 19]) on Hadoop MapReduce and made his way through a number of DB textbooks. He remembers a sentence from one of his professors saying "full-table-scans are bad; indexes are good"[1]. He finds a paper that shows how to create a so-called *trojan index* [12], i.e. an index that may be used with Hadoop MapReduce and yet does not modify the underlying Hadoop MapReduce and HDFS engines. Therefore, Bob decides to create a trojan index on sourceIP before running his MapReduce jobs. However, using trojan indexes raises two other problems:

**(1.)** *Expensive index creation.* The time to create the trojan index on sourceIP (or any other attribute) is very long, actually it is much longer than running a scan-based query on all his data in the first place. If Bob's MapReduce jobs use that index only a few times, the index creation costs will never be amortized. So, why would Bob create such an expensive index in the first place?

[1] The professor is aware that for some situations the opposite is true.

**(2.)** *Which attribute to index?* Even if Bob amortizes index creation costs by running a dozen queries filtering data on sourceIP: the trojan index will only help for that particular attribute. However, the trojan index will not help when searching for other attributes, e.g. visitDate or adRevenue. So, which attribute should Bob use to create the index?

One day in autumn 2011, Bob reads about another idea [21] where some researchers looked at ways to improve vertical partitioning in Hadoop. That work considered physical data layouts, e.g. row, column, PAX [2], and vertical layouts. A major problem with those existing layouts was that they are good for one particular type of workload (e.g. requesting many attributes), but bad for others (e.g. requesting only few attributes). Would there be a neat way to use multiple data layouts at the same time? The researchers of [21] realized that HDFS keeps three (or more) copies of all data anyway. These copies were originally kept for failover only. Hence, all copies used exactly the same physical data layout. Therefore they decided to change HDFS to store each copy in a *different* vertical layout. The different layouts were computed by an algorithm. As all layout transformation was done per HDFS block, the failover properties of HDFS and Hadoop MapReduce were not affected at all. At the same time, I/O times improved.

Bob thinks that this looks interesting. However, would that solve his indexing problem?

This is where the story begins.

## 1.1 Idea

We propose HAIL (Hadoop Aggressive Indexing Library): an enhancement of HDFS and Hadoop MapReduce. HAIL keeps the existing physical replicas of an HDFS block in different sort orders and with different clustered indexes. Hence, for a default replication factor of three at least three different sort orders and indexes are available for MapReduce job processing. Thus, the likelihood to find a suitable index increases and hence the runtime for a workload improves. We modify the upload pipeline of HDFS to already create those indexes while uploading data to HDFS. Therefore, no additional read of the data is required. No additional MapReduce jobs are required to create those indexes. The decision on the indexes to create can either be done by a user through a configuration file or by a physical design algorithm. HAIL typically improves both the upload times (even if index creation is included) and the MapReduce job execution times. Therefore, HAIL provides a win-win situation over Hadoop MapReduce and even over Hadoop++.

It is worth noting that even if in this paper we illustrate the benefits of HAIL via a web log data processing example, many more data analytics applications (such as OLAP and scientific applications) can benefit from using HAIL. However, discussing each of these applications is beyond the scope of this paper.

## 1.2 Research Challenges and Questions

We face a number of key challenges:

**(1.)** How can we change HDFS to create indexes already when uploading files from outside into HDFS? How can we support different sort orders and indexes for different replicas? Which changes to the HDFS upload pipeline need to be done to make this efficient? What happens to the involved checksum mechanism of HDFS? How can we teach the HDFS namenode to distinguish the different replicas and keep track of the different indexes?

**(2.)** How can we change Hadoop MapReduce to fully exploit the different sort orders and indexes at query time? How much do we need to change existing MapReduce jobs? How can we change Hadoop MapReduce to schedule tasks to replicas having the appropriate index? What happens to Hadoop MapReduce failover? How will Hadoop MapReduce change from the user's perspective?

## 1.3 Contributions and Answers

**(1.)** We show how to effectively piggy-back sorting and index creation on the existing HDFS upload pipeline. In fact, the HAIL upload pipeline is so effective when compared to HDFS that the additional overhead for sorting and index creation is hardly noticeable in the overall process. HAIL even allows us to create more than three indexes at reasonable costs. Our approach also benefits from the fact that Hadoop is only used for appends: there are no updates. Therefore, once a block is full it will never be changed again. We will first give an overview of our system and its benefits (Section 2) and then explain the differences between the HAIL and Hadoop upload pipeline in more detail (Section 3).

**(2.)** We show how to effectively change the Hadoop MapReduce pipeline to exploit HAIL indexes. We do this in a minimally invasive manner, only changing the RecordReader and a few UDFs. Moreover, we show how to allow users (or query optimizer) to easily exploit our indexes (Section 4).

**(3.)** We present an extensive experimental comparison of HAIL with Hadoop and Hadoop++ [12]. We use six different clusters including physical and virtual EC2 clusters of up to 100 nodes. A series of scalability experiments (#indexes, #replicas, cluster scale-up, cluster scale-out, and failover) with different datasets demonstrates the superiority of HAIL (Section 6.3).

**(4.)** We show that the Hadoop MapReduce framework incurs a very high scheduling overhead for short running jobs. As a result, these jobs cannot fully benefit from using clustered indexes (Section 6.4). HAIL reduces this overhead significantly using a novel splitting policy to partition data at query time. This splitting policy together with clustered indexes allows HAIL to run up to 68x faster than Hadoop (Section 6.5).

# 2. OVERVIEW

We start with an overview contrasting HAIL with the current HDFS and Hadoop MapReduce. At the same time we introduce the terminology used in the reminder of this paper. For a more detailed discussion on differences of HAIL to related work see Section 5.

Let's return to Bob again. How can Bob analyze his log file using the different systems?

## 2.1 Hadoop and HDFS

With the existing HDFS and Hadoop MapReduce stack, Bob starts by uploading his log file to HDFS using the *HDFS client*. HDFS then partitions the file into logical *HDFS blocks* using a constant block size (the HDFS default is 64MB). Each HDFS block is then physically stored three times (assuming the default replication factor). Each physical copy of a block is called a *replica*. Each replica will sit on a different *datanode*. Therefore, at least two datanode failures may be survived by HDFS. Note that HDFS keeps information on the different replicas for an HDFS block in a central *namenode* directory.

After uploading his log file to HDFS, Bob may run an actual MapReduce job. Bob invokes Hadoop MapReduce through a Hadoop MapReduce *JobClient*, which sends his *MapReduce job* to a central node termed *JobTracker*. The MapReduce job consists of several *tasks*. A task is executed on a subset of the input file, typically an HDFS block[2]. The JobTracker assigns each task to a different *TaskTracker* — which typically runs on the same machine as an HDFS datanode. Each datanode will then read its subset of the input file, i.e. a set of HDFS blocks, and feed that data into the *MapReduce processing pipeline* which usually consists of a Map, Shuffle, and a Reduce Phase (see [10, 12, 11] for a detailed description). As soon as all results have been written to HDFS, Bob will be informed that the result sets are available. The execution time of the MapReduce job is heavily influenced by the size of the initial input because Hadoop MapReduce reads the input file(s) entirely for each MapReduce job.

[2] Actually it is a *split*. The difference does not matter here. We will get back to this in Section 4.2.

## 2.2 HAIL

In HAIL Bob analyzes his log file as follows. He starts by uploading his log file to HAIL using the *HAIL client*. In contrast to the HDFS client, the HAIL client analyzes the input data for each HDFS block, converts each HDFS block directly to binary PAX layout [2] and sends it to three datanodes. Then all datanodes sort the data contained in that block in parallel using a different sort order — as manually specified by Bob in a configuration file or as computed by a physical design algorithm. All sorting and index creation happens in main memory. This is feasible as the block size is typically between 64MB (default) and 1GB. This easily fits into the main memory of most machines. In addition, in HAIL, each datanode creates a different clustered index for each data block and stores it with the sorted data.

After uploading his log file to HAIL, Bob may run his MapReduce job exploiting the indexes created by HAIL. As before, Bob invokes Hadoop MapReduce through a JobClient which sends his MapReduce job to the JobTracker. However, his MapReduce job is slightly changed to exploit the indexes available on the different replicas in HAIL. For instance, assume that an HDFS block has three replicas with clustered indexes on visitDate, adRevenue, and sourceIP. Depending on the index required, Hadoop MapReduce running on top of HAIL will use the replica with the suitable index. If Bob has a MapReduce job filtering on visitDate, HAIL will use the replica having the clustered index on visitDate. If Bob is filtering on sourceIP, HAIL will use the replica having the clustered index on sourceIP and so on. To provide failover and load balancing, HAIL may sometimes fall back to using a replica without a useful index for some of the blocks, i.e. it will fall back to standard Hadoop scanning. However — even factoring this in — Bob's queries will on average run much faster.

## 2.3 HAIL Benefits

**(1.)** HAIL often improves both upload *and* query times. The upload is dramatically faster than Hadoop++ and often faster (or only slightly slower) than with the standard Hadoop even though we (i) convert the input file into binary PAX, (ii) create a series of different sort orders, and (iii) create multiple clustered indexes. From the user-side this provides a win-win situation: there is no noticeable punishment for upload. For querying, the user can only win: if our indexes cannot help, we will fall back to standard Hadoop scanning. If the indexes can help, query runtimes will improve.

*Why don't we have high costs at upload time?* We basically exploit the unused CPU ticks which are not used by standard HDFS. As the standard HDFS upload pipeline is I/O-bound, the effort for our sorting and index creation in the HAIL upload pipeline is hardly noticeable. In addition, as we already parse data to binary while uploading, we often benefit from a smaller binary representation triggering less network and disk I/O.

**(2.)** We do not change the failover properties of Hadoop.

*Why is failover not affected?* All data stays on the same *logical* HDFS block. We just change the *physical* representation of each replica of an HDFS block. Therefore, from each replica we may recover the logical HDFS block.

**(3.)** HAIL works with existing MapReduce jobs incurring only minimal changes to those jobs.

*Why does this work?* We allow Bob to annotate his existing jobs with selections and projections. Those annotations are then considered by HAIL to pick the right index. Like that, for Bob the changes to his MapReduce jobs are minimal.

# 3. THE HAIL UPLOAD PIPELINE

Let's take a look at the HAIL upload process in more detail. As you will see, there are some surprising challenges to tackle when changing HDFS to create different indexes per replica. Figure 1 shows the data flow when Bob is uploading a file to HAIL.

## 3.1 Data Transformation

**In HDFS**, for each block, the client contacts the namenode to obtain the list of datanodes that should store the block replicas. Then, the client sends the original block to the first datanode, which forwards this to the second datanode and so on.

**In HAIL**, the HAIL client preprocesses the file based on content to consider end of lines ① in Figure 1. We parse the contents into rows by searching for end of line symbols and never split a row between two blocks. This is in contrast to standard HDFS which splits a file into HDFS blocks after a constant number of bytes. For each block the HAIL client parses each row according to the schema specified by the user[3]. If HAIL encounters a row that does not match the given schema (i.e., a bad record), it separates this record into a special part of the data block. HAIL then converts all data blocks to a binary PAX representation ②. The HAIL client also collects metadata information from each data block (such as the data schema) and creates a block header (*Block Metadata*) for each data block ②.

If we piggy-backed naively on this existing HDFS upload pipeline by first storing the original block data as done in Hadoop and then converting it to binary PAX layout in a second step, we would have to re-read and then re-write each block to create the index. This would trigger one extra write and one extra read *for each replica*, e.g. for an input file of a 100GB we would have to pay 600GB extra I/O on the cluster. In fact, one of our first prototypes followed the naive approach; this lead to very long upload times. In contrast, HAIL has an important benefit: we do not have to pay any of that extra I/O. However, to achieve this dramatic improvement, we have to make non-trivial changes in the standard Hadoop upload pipeline.

## 3.2 Upload Pipeline

To understand the HAIL upload pipeline, we first have to analyze the existing HDFS pipeline in more detail.

**In HDFS**, while uploading a block, the data is further partitioned into *chunks* of constant size 512B. Chunks are collected into *packets*. A packet is a sequence of chunks plus a checksum for each of the chunks. In addition some metadata is kept. In total a packet has a size of up to 64KB. Immediately before sending the data over the network, each HDFS block is converted to a sequence of packets. On disk, HDFS keeps, for each replica, a separate file containing checksums for all of its chunks. Hence, for each replica two files are created on local disk: one file with the actual data and one file with its checksums. These checksums are reused by HDFS whenever data is send over the network, e.g. if the data is read by a remote datanode. The HDFS client (CL) sends the first packet of the block to the first datanode in the upload pipeline (DN$_1$). DN$_1$ splits the packet into two parts: the first contains the actual chunk data, the second contains the checksums for those chunks. Then DN$_1$ flushes the chunk data to a file on local disk. The checksums are flushed to an extra file. In parallel DN$_1$ forwards the packet to DN$_2$ which splits and flushes the data like DN$_1$ and in turn forwards the packet to DN$_3$ which splits and flushes the data as well. Yet, only DN$_3$ verifies the checksum for each chunk. If the recomputed checksums for each chunk of a packet matches the received checksums, DN$_3$ acknowledges the packet back to DN$_2$, which acknowledges back to DN$_1$. Finally, DN$_1$ acknowledges back to CL. Each datanode also appends its ID to the ACK. Like that only one of the datanodes (the last in the chain, here DN$_3$ as the replication factor is three) has to verify the checksums. DN$_2$ believes DN$_3$, DN$_1$ believes DN$_2$, and CL believes DN$_1$.

[3]Alternatively, HAIL may suggest an appropriate schema to users.

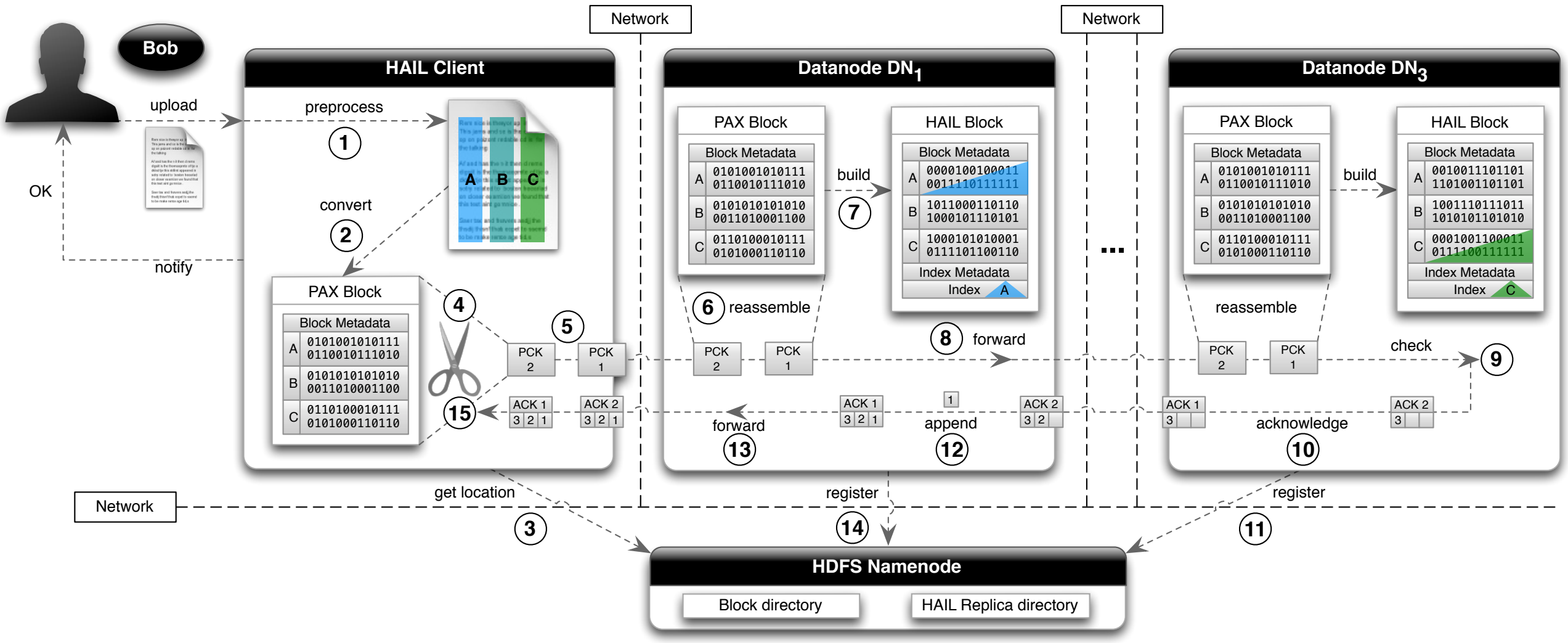

**Figure 1: The HAIL upload pipeline**

If any CL or DN$_i$ receives ACKs in the wrong order, the upload has failed. The idea of sending multiple packets from CL is to hide the roundtrip latencies of the individual packets. Creating this chain of ACKs also has the benefit that CL only receives a single ACK for each packet and not three. Notice that HDFS provides this checksum mechanism on top of the existing TCP/IP checksum mechanism (which has weaker correctness guarantees than HDFS). **In HAIL**, in order to reuse as much of the existing HDFS pipeline and yet to make this efficient, we need to perform the following changes. As before the HAIL client gets the list of datanodes to use for this block from the HDFS namenode (3). But rather than sending the original input, CL creates the PAX block, cuts it into packets (4) and sends it to DN$_1$ (5). Whenever a datanode DN$_1$–DN$_3$ receives a packet, it does *neither* flush its data *nor* its checksums to disk. Still DN$_1$ and DN$_2$ immediately forward the packet to the next datanode as before (8). DN$_3$ will verify the checksum of the chunks for the received PAX block (9) and acknowledge the packet back to DN$_2$ (10). This means the semantics of an ACK for a packet of a block are changed from "packet received, validated, and flushed" to "packet received and validated". We do neither flush the chunks nor its checksums to disk as we first have to sort the entire block according to the desired sort key. We assemble the block from all packets in main memory (6). This is realistic in practice for most modern servers, as main memories tend to be beyond 10GB for any modern server. Typically, the size of a block is between 64MB (default) and 1GB. This means that for the default size we could keep about 150 blocks in main memory at the same time.

In parallel to forwarding and reassembling packets, each datanode sorts the data, creates indexes, and forms a *HAIL Block* (7), (see Section 3.5). As part of this process, each datanode also adds *Index Metadata* information to each data block in order to specify the index it created for this block. Each datanode (e.g., DN$_1$) typically sorts the data inside a block in a different sort order. It is worth noting that having different sort orders across replicas does not impact fault-tolerance as all data is reorganized *inside the same block only* — data is *not* reorganized *across* blocks. As soon as a datanode has completed sorting and creating its index, it will recompute checksums for each chunk of a block. Notice that checksums will differ on each replica, as different sort orders and indexes are used. Hence, each datanode has to compute its own checksums. Then each datanode flushes the chunks and newly computed checksums to two separate files on local disk as before. For DN$_3$, once all chunks and checksums have been flushed to disk, DN$_3$ will acknowledge the last packet of the block back to DN$_2$ (10). After that DN$_3$ will inform the HDFS namenode about its new replica including its HAIL block size, the created indexes, and the sort order (11) (see Section 3.3). Datanodes DN$_2$ and DN$_1$ append their ID to each ACK (12). Then they forward each ACK back in the chain (13). DN$_2$ and DN$_1$ will forward the last ACK of the block only if all chunks and checksums have been flushed to their disks. After that DN$_2$ and DN$_1$ individually inform the HDFS namenode (14). The HAIL client also checks that all ACKs arrive in the right order (15).

To keep track of the different sort orders it is important to change the HDFS namenode as well. We discuss how the namenode maintains the sort order for each data block replica in Section 3.3.

## 3.3 HDFS Namenode Extensions

**In HDFS,** the central namenode keeps a directory `Dir_block` of blocks, i.e. a mapping blockID $\mapsto$ Set Of DataNodes. This directory is required by any operation retrieving blocks from HDFS. Hadoop MapReduce exploits `Dir_block` for scheduling. In Hadoop MapReduce whenever a block needs to be assigned to a worker in the map phase, the scheduler looks up `Dir_block` in the HDFS namenode to retrieve the list of datanodes having a replica of that block. Then the Hadoop MapReduce scheduler will try to schedule map tasks on those datanodes if possible. Unfortunately, the HDFS namenode does not differentiate the replicas any further w.r.t. their physical layouts. HDFS was simply not designed for this. From the point of view of the namenode all replicas are byte-equivalent and have the same size.

**In HAIL,** we need to allow Hadoop MapReduce to change the scheduling process to schedule map tasks close to replicas having a suitable index — otherwise Hadoop MapReduce would pick indexes randomly. Hence we have to enrich the HDFS namenode to keep additional information about the available indexes. We do this by keeping an additional directory `Dir_rep` mapping (blockID, datanode) $\mapsto$ HAILBlockReplicaInfo. An instance of HAILBlockReplicaInfo contains detailed information about the types of available indexes for a replica, i.e. indexing key, index type, size, start offsets, and so on. As before, Hadoop MapReduce looks up `Dir_block` to retrieve the list of datanodes having a replica for a given block. However, in addition, HAIL looks up the main memory `Dir_rep` to obtain the detailed HAILBlockReplicaInfo for each replica, i.e. one main memory lookup for each replica. HAILBlockReplicaInfo is then exploited by HAIL to change the scheduling strategy of Hadoop MapReduce (we will discuss this in detail in Section 4).

## 3.4 Which Attributes to Index?

Bob is happy doing his web-log analysis using HAIL. The web logs contain just a few attributes and he configures HAIL manually to simply create indexes on all of them. But what if Bob's dataset contains more attributes than the number of replicas? Picking the right indexes is not easy in such cases. Over the years, databases researchers have proposed several algorithms [9, 4, 6, 1] to select the indexes to create, given a query workload. However, these index selection algorithms do not take into account default data replication (which is the case for HDFS).

Our recently proposed Trojan Layouts algorithm [21] overcomes this problem for vertical partitioning, i.e. it respects data block replication in HDFS and creates different physical vertical layouts for the different replicas. However, the Trojan Layouts algorithm is strictly limited to vertical partitioning. In contrast, for HAIL it would be interesting to have an algorithm that can propose different clustered indexes for different replicas. We believe that [21] can be extended to compute these indexes. Actually we believe that it can even be extended to compute *both* vertical partitions *and* indexes for different replicas simultaneously. However, this research leads way beyond the scope of this paper and we will investigate it as part of future work.

## 3.5 Indexing Pipeline

**Why Clustered Indexes?** An interesting question is why we focus on clustered indexes? We require an index structure that is cheap to *create in main memory*, cheap to *write to disk*, and cheap to *query from disk*. We tried a number of indexes in the beginning of the project — including coarse-granular indexes and unclustered indexes. After some experimentation we quickly discovered that sorting and index creation in main memory is so fast that techniques like partial or coarse-granular sorting do not pay off for HAIL. Whether you pay three or two seconds for sorting and indexing per block during upload is hardly noticeable in the overall upload process of HDFS. In addition, a major problem with unclustered indexes is that they are only competitive for very selective queries as they may trigger considerable random I/O for non-selective index traversals. In contrast, clustered indexes do not have that problem. Whatever the selectivity, we will read the clustered index and scan the qualifying blocks. Hence, even for very low selectivities the only overhead over a scan is the initial index node traversal — which is very cheap in comparison to the scan and especially in comparison to an *unclustered* index traversal. Moreover, as unclustered indexes are dense by definition, they require considerably more space on disk and require more write I/O than a sparse clustered index. Thus, using unclustered indexes would severely affect upload times. Yet an interesting direction for future work would be to extend HAIL to support additional indexes which might boost the performance of our system even further including bitmap indexes for low cardinality domains or inverted lists for untyped or bad records, i.e. records not obeying a specific schema.

**Column Index Structure.** All sorting and index creation happens in main memory; this is a realistic assumption as the HDFS block size is small (typically below 1GB). Hence, on each data node several blocks may be indexed in parallel entirely in main memory. Let's assume we want to create a clustered index on sourceIP. We first sort the data using sourceIP as the sort key. We need to preserve the sort order among all columns in the PAX-block. Therefore, we build a sort index to reorganize all other columns as explained above. After that we create a *sparse clustered $B^+$-tree*, which has a single large root directory. It logically divides the data of attribute sourceIP into *partitions* consisting of 1,024 values and has child pointers to their start offsets (see Figure 2). All but the first child pointer are implicitly defined as all leaves are contiguous on disk and can be reached by simply multiplying the leaf size with the leaf ID. The structure has some similarities with a CSB$^+$-tree [29], but our index is different in that we keep all leaves contiguous on disk rather than in main memory. In addition, we keep a single directory since index lookups are dominated by disk seeks.

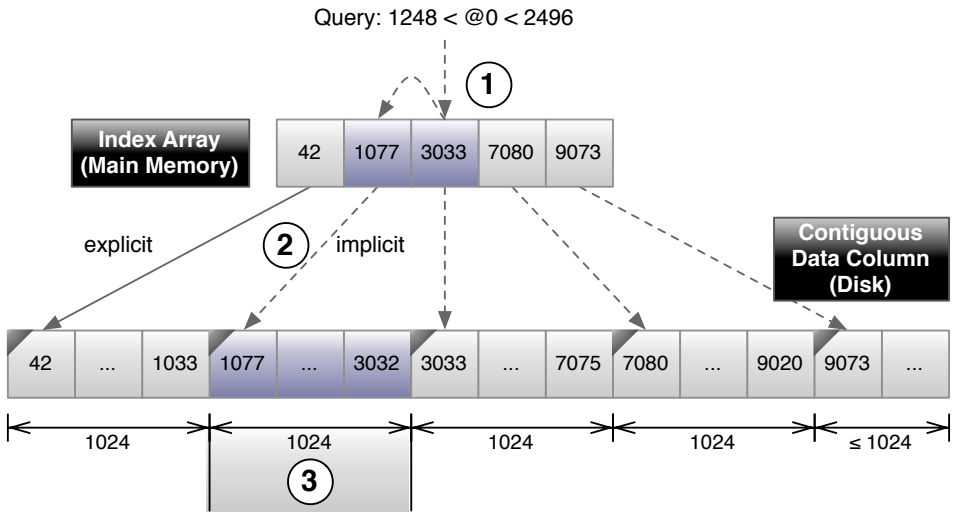

**Figure 2: HAIL data column index**

Why not a multi-level tree? For instance, assume that the input dataset has 10 attributes of a fixed-size 4B each, i.e. each row occupies 40B. Hence each 256MB block stores 6.7 million rows. Each attribute occupies 25.6MB. If we assume a page size of 4KB this is 25.6MB/4KB=6,554 pages per attribute. Therefore the root node has 6,554 entries of 4B each, i.e. in total 25.6KB. This represents an overhead of 0.01% over the data block size[4]. Let's assume a realistic hard disk transfer rate of 100MB/sec. Hence the root node may be read within $\sim$0.3ms plus the initial seek of 5ms, i.e. 5.3ms total read time. In contrast, a two-level index access would trigger an extra seek, i.e. in total 5+5=10ms seek costs plus data transfer.

How big does the HDFS block have to be to justify a multilevel index? We compute the maximum index size as 100MB/sec/5ms = 500KB. This index corresponds to 125,000 index entries, hence 512MB per attribute and therefore 5GB HDFS block size. Only for larger HDFS blocks a multi-level index would pay off. We did not implement multi-level trees as block sizes are usually below 1GB. In contrast to a standard B$^+$-tree range query we already determine the first and the last partition of the data to read in main memory ① & ② to avoid post-filtering the entire range ③.

**Accessing Variable-size Attributes.** Accessing variable-size attributes is different from accessing fixed-size attributes like sourceIP in that we cannot simply calculate the offsets of partitions anymore. In general, we store variable-sized attributes as a sequence of zero-terminated values. Whenever we index a block, we also create additional lists of offsets for all variable-size attributes and store them in front of their corresponding data. Notice that we only need to store one offset for each logical partition. Hence we only store every $n$-th offset, where $n$ is the number of values in each partition. With this minor modification our index supports tuple reconstruction for variable-size attributes. For instance, assume a query filtering on fixed-size sourceIP and projecting to variable-length URL. The index on sourceIP returns a number of qualifying rowIDs. For these rowIDs we have to retrieve their variable-length URL values. We simply do this by looking up the offsets to the next partition in main memory. Assume we need to retrieve the URL for rowID=43,425 and each partition has 1,024 entries. Then we scan the partition $\lfloor 43{,}425/1{,}024 \rfloor = 42$ entirely from disk. The overheads for scanning that partition over the initial random I/O are small. Then, in main memory we post-filter the partition to retrieve the URL for rowID 43,425.

[4]The overhead of an unclustered index would be about 10% to 20% over the data block size.

# 4. THE HAIL QUERY PIPELINE

We now focus on how Bob builds his MapReduce jobs and the way HAIL executes MapReduce jobs. From Bob's perspective, we will see in Section 4.1 that Bob has to write his MapReduce jobs (almost) as before and run them exactly as when using Hadoop MapReduce. From the system perspective, we first analyze the standard Hadoop MapReduce pipeline in Section 4.2 and then see how HAIL executes MapReduce jobs in Section 4.3. We will see that HAIL requires only small changes in the Hadoop MapReduce framework, which makes HAIL easy to integrate into newer Hadoop versions. Figure 3 shows the query pipeline when Bob runs a MapReduce job on HAIL.

## 4.1 Bob's Perspective

**In Hadoop MapReduce**, Bob writes a MapReduce job, including a job configuration class, a map function, and a reduce function.

**In HAIL**, the MapReduce job remains the same (see ① and ② in Figure 3), but with three tiny changes:

(1) Bob specifies the `HailInputFormat` (which uses a `HailRecordReader`[5] internally) in the main class of the MapReduce job. By doing this, Bob enables his MapReduce job to read HAIL Blocks (see Section 3.2).

(2) Bob annotates his map function to specify the selection predicate and the projected attributes required by his MapReduce job[6]. For example, assume that Bob wants to write a MapReduce job that performs the following SQL query (example from Introduction):

```
SELECT sourceIP
FROM UserVisits WHERE visitDate
BETWEEN '1999-01-01' AND '2000-01-01';
```

To execute this query in HAIL, Bob adds to his map function a `HailQuery` annotation as follows:

```
@HailQuery(filter="@3 between(1999-01-01,
  2000-01-01)", projection={@1})
void map(Text key, Text v) { ... }
```

Where the literal `@3` in the `filter` value and the literal `@1` in the `projection` value denote the attribute position in the `UserVisits` records. In this example the third attribute (i.e. `@3`) is `visitDate` and the first attribute (i.e. `@1`) is `sourceIP`. By annotating his map function as mentioned above, Bob indicates that he wants to receive in the map function only the projected attribute values of those tuples qualifying the specified selection predicate. In case Bob does not specify filter predicates, MapReduce will perform a full scan on the HAIL Blocks as in standard Hadoop. At query time, if the `HailQuery` annotation is set, HAIL checks (using the *Index Metadata* of a data block) whether an index exists on the filter attribute. Using such an index allows us to speed up the execution of a map task. HAIL also uses the *Block Metadata* to determine the schema of a data block. This allows HAIL to read the attributes specified in the filter and projection parameters correctly.

(3) Bob uses a `HailRecord` object as input value in the map function. This allows Bob to directly read the projected attributes without splitting the record into attributes as he would do it in the standard Hadoop MapReduce. For example, using standard Hadoop MapReduce Bob would write the following map function to perform the above SQL query:

MAP FUNCTION FOR HADOOP MAPREDUCE (PSEUDO-CODE):

```
void map(Text key, Text v) {
   String[] attributes = v.toString().split(",");
   if (DateUtils.isBetween(attributes[2],
        "1999-01-01", "2000-01-01"))
      output(attributes[0], null);
}
```

Using HAIL Bob writes the following map function:

MAP FUNCTION FOR HAIL:

```
void map(Text key, HailRecord v) {
   output(v.getInt(1), null);
}
```

Notice that Bob now does not have to filter out the incoming records, because this is automatically handled by HAIL via the `HailQuery` annotation (as mentioned earlier). This new map function as well as the annotation is illustrated in Figure 3.

## 4.2 Hadoop Perspective

**In Hadoop MapReduce**, when Bob submits a MapReduce job a *JobClient* instance is created. The main goal of the JobClient instance is to copy all the resources needed to run the MapReduce job (e.g. metadata and job class files). But also, the JobClient fetches all the block metadata (`BlockLocation[]`) of the input dataset. Then, the JobClient logically breaks the *input* into smaller pieces called *input splits* (*split phase* in Figure 3) using the `InputFormat`-UDF. By default, the JobClient computes input splits such that each input split maps to a distinct HDFS block. Notice that an input split defines the input data of a map task. On the other hand, a data block is a horizontal partition of a dataset stored in HDFS (see Section 3.1 for details on how HDFS stores datasets). For scheduling purposes, the JobClient retrieves for each input split all datanode locations having a replica of that block. This is done by calling the `getHosts` method of each `BlockLocation`. For instance in Figure 3, block$_{42}$ is stored on datanodes `DN3`, `DN5`, and `DN7`, and hence these datanodes are the *split locations* for split$_{42}$.

After this split phase, the JobClient submits the job to the *JobTracker* with a set of splits to process ③. Among other operations, the JobTracker creates a *map task* for each input split. Then, for each map task, the JobTracker decides on which computing node to schedule the map task, using the split locations ④. This decision is based on data-locality and availability [10]. After this, the JobTracker allocates the map task to the *TaskTracker* (which performs map and reduce tasks) running on that computing node ⑤.

Only then, the map task can start processing its input split. The map task uses a *RecordReader* UDF in order to read its input data block$_i$ from the closest datanode ⑥. Interestingly, it is the *local HDFS client* running on the node where the map task is running that decides from which datanode a map task will read its input — and *not* the Hadoop MapReduce scheduler. This is done when the RecordReader asks for the input stream pointing to block$_i$. It is worth noting that the HDFS client chooses a datanode from the set of all datanodes storing a replica of $block_{42}$ (via the `getHosts` method) rather than from the locations given by the input split. This means that a map task might eventually end up reading its input data from a remote node even though it is available locally. Once the input stream is opened, the RecordReader breaks $block_{42}$ into records and makes a call to the map function for each record. Assuming that the MapReduce job consists of a map phase only, the map task then writes its output back to HDFS (7). See [12, 33, 11] for more details on the Hadoop MapReduce execution pipeline.

[5]Which is also called *itemize* UDF in Hadoop++ [12].

[6]Alternatively, HAIL allows Bob to specify the selection predicate and the projected attributes in the job configuration class.

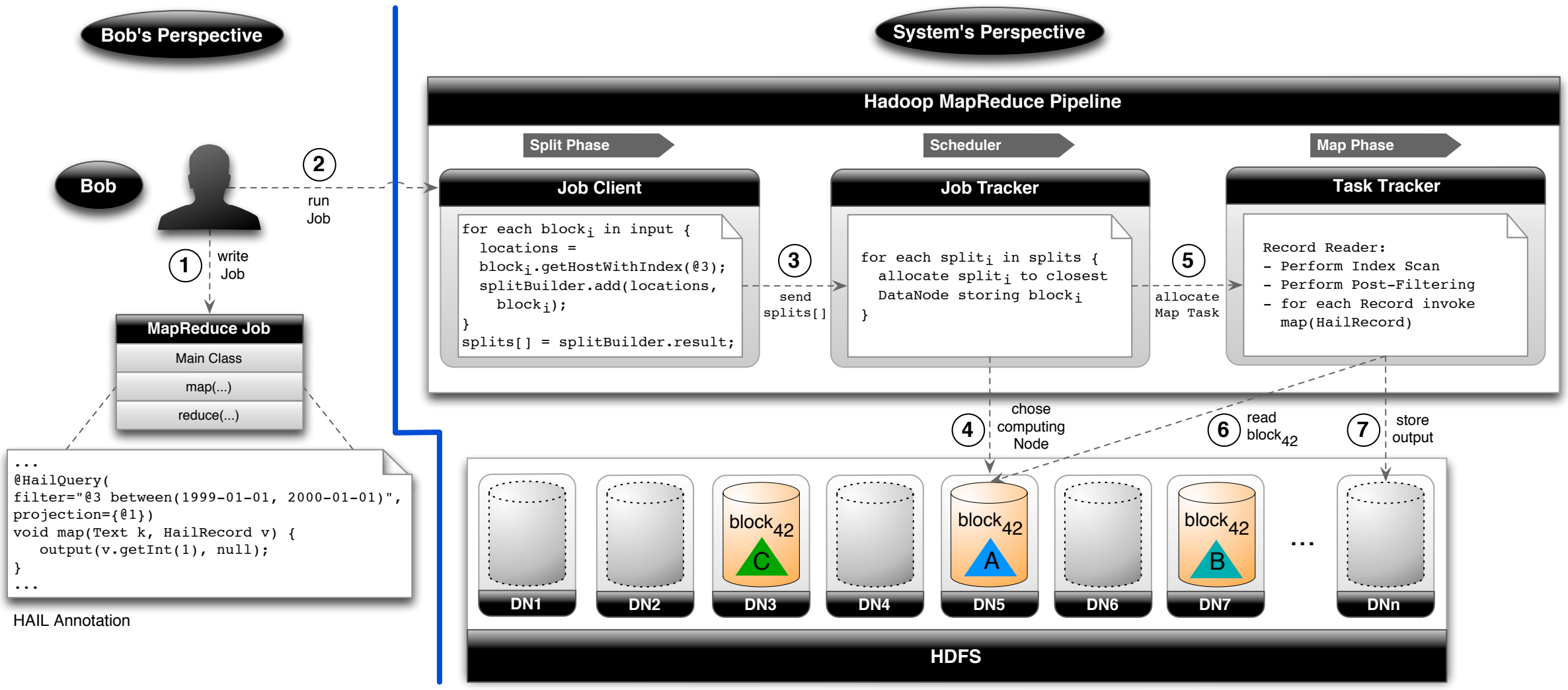

**Figure 3: The HAIL query pipeline (with default Hadoop scheduling)**

## 4.3 HAIL Perspective

**In HAIL**, it is crucial to be non-intrusive to the standard Hadoop execution pipeline so that users run MapReduce jobs exactly as before. However, supporting per-replica indexes in an efficient way and without significant changes in the standard execution pipeline is challenging for several reasons. First, the JobClient cannot simply create input splits based only on the default block size as each block replica has a different size (because of indexes). Second, the JobTracker can no longer schedule map tasks based on data-locality and nodes availability only. The JobTracker now has to consider the existing indexes on each block replica. Third, the RecordReader has to perform either index access or full scan of data blocks without any interaction with users. Fourth, the HDFS client cannot anymore open an input stream to a given block based on data-locality and nodes availability only (it has to consider existing indexes as well). HAIL overcomes these issues by mainly providing two UDFs: the `HailInputFormat` and the `HailRecordReader`. Using UDFs, we allow HAIL to be easy to integrate into newer versions of Hadoop MapReduce.

In contrast to the Hadoop MapReduce `InputFormat`, the `HailInputFormat` uses a more elaborate splitting policy, called `HailSplitting`. The overall idea of `HailSplitting` is to map one input split to several data blocks whenever a MapReduce job performs an index scan over its input. This allows HAIL to reduce the number of map tasks to process and hence to reduce the aggregated cost of initializing and finalizing map tasks. The reader might think that using several blocks per input split may significantly impact failover. However, this is not true since jobs performing an index scan are relatively short running jobs (in the order of a few seconds). Therefore, the probability that one node fails in this period of time is very low [28]. In case a node fails in this period of time, HAIL simply reschedules the failed map tasks, which results only in a few seconds overhead anyways. Optionally, HAIL could apply the checkpointing techniques proposed in [28] in order to improve failover. We will study these interesting aspects in a future work. Notice that for those MapReduce jobs performing a full scan, `HailSplitting` still uses the default Hadoop splitting, i.e., it creates an input split for each data block. Hence, failover for these jobs is not changed at all.

To improve data locality, `HailSplitting` first clusters the blocks of the input of an incoming MapReduce job by locality. As a result of this process, `HailSplitting` produces as many collections of blocks as there are datanodes storing at least one block of the given input. Then, for each collection of blocks, `HailSplitting` creates as many input splits as map slots each TaskTracker has. HAIL creates a map task per resulting input split and schedules these map tasks to the replicas having the matching index. For example in Figure 3, `DN5` has the matching clustered index to process $split_{42}$, hence the JobTracker schedules map task for $split_{42}$ to `DN5` (or close to it). The reader might think that performance could be negatively impacted in case that data locality is not achieved for several map tasks. However, fetching small parts of blocks through the network (which is the case when using index scan) is negligible [21]. Moreover, one can significantly improve data locality by simply using an adequate scheduling policy (e.g. the Delay Scheduler [34]). If no relevant index exists, HAIL scheduling falls back to standard Hadoop scheduling by optimizing data locality only.

The `HailRecordReader` is responsible for retrieving the records that satisfy the selection predicate of MapReduce jobs (as illustrated in the MapReduce Pipeline of Figure 3). Those records are then passed to the map function. For example in Bob's query of Section 4.1, we need to find all records having `visitDate between(1999-01-01, 2000-01-01)`. To do so, we first open an input stream to the block having the required index. For this, HAIL instructs the local HDFS Client to use the newly introduced `getHostsWithIndex` method of each `BlockLocation` so as to choose the closest datanode with the required index. Once that input stream has been opened, we use the information about selection predicates and attribute projections from the `HailQuery` annotation or from the job configuration file. When performing an index-scan, we read the index entirely into main memory (typically a few KB) to perform an index lookup. This also implies reading the qualifying block parts from disk into main memory and post-filtering records (see Section 3.5). Then, we reconstruct the projected attributes of qualifying tuples from PAX to row layout. In case that no projection was specified by users, we then reconstruct all attributes. Finally, we make a call to the map function for each qualifying tuple. For bad records (see Section 3.1), HAIL passes them directly to the map function, which in turn has to deal with them (just like in standard Hadoop MapReduce). For this, the `HailRecord` provides a flag to indicate bad records. If full scan is used, the `HailRecordReader` still applies the selection predicate and performs tuple reconstruction.

# 5. RELATED WORK

The closest work to HAIL is Hadoop++ [12], which creates a logical block-level index. However, Hadoop++ can only create this so-called *trojan index* per *logical* HDFS block rather than per *physical* replica as in HAIL. In addition, index creation in Hadoop++ is very expensive, as after uploading the input file to HDFS, Hadoop++ uses an additional MapReduce job to convert the data to binary format and to create the trojan index. We collect considerable evidence on this in the experiments.

In another related work [15], researchers from Twitter proposed a full text indexing technique for improving Hadoop performance. However, this indexing technique is not well suited for analytical and exploratory queries as considered in this paper. This is because full text indexes are only suitable for highly selective queries as already concluded in [15]. Nevertheless, we ran micro-benchmarks for upload and index creation times. We observed that [15] required $2,088$ seconds to only create a full-text index on 20GB, while HAIL takes $1,600$ seconds to both upload and index 200GB.

Recently, CoHadoop [13] improved the co-partitioning features of Hadoop++. However, CoHadoop did not improve any of the indexing features of Hadoop++, which is the focus of HAIL. Manimal [5] proposed to analyze MapReduce jobs to determine filter conditions. Then, MapReduce jobs are rewritten to match an existing index. Again, Manimal only considers logical indexes on the block level and not per replica indexes. Manimal's MapReduce job code analysis could be combined with our system. Cloud indexing [7] creates a P2P overlay on a network of virtual machines, which is very similar to established P2P systems like Chord [31]. The main idea is to setup an extra indexing service on top of the underlying cloud data service adding extra resources like main memory. That is not what we propose: our idea is to integrate indexing with an existing service: in our case HDFS. In terms of data layouts we used PAX [2], which was originally invented for cache-conscious processing but adapted by a number of other people in the context of MapReduce [8, 14]. In our previous work [21], we showed how to improve PAX by computing different layouts on the different replicas. However, in that work we did not consider indexing. This paper fills this gap.

To the best of our knowledge, this is the first work that aims at pushing indexing to the extreme at low index creation cost.

# 6. EXPERIMENTS

Let's get back to Bob again and his initial question: *will HAIL solve his indexing problem efficiently?* Overall, we need to answer the following questions experimentally:

**(1.)** What is the performance of HAIL at upload time? What is the impact of HAIL indexing in the upload pipeline? How many indexes can we create in the time the standard HDFS uploads the data? How does hardware performance affect HAIL upload on a cluster of nodes? How well does HAIL scale-out on large clusters? (We answer these questions in Section 6.3).

**(2.)** What is the performance of HAIL at query time? How much do query sequences benefit from HAIL? How much do RecordReader times benefit from HAIL? How does query selectivity affect HAIL? How do failing nodes affect HAIL? (We answer these questions in Section 6.4). How does `HailSplitting` improve end-to-end job runtimes? (We answer this question in Section 6.5).

## 6.1 Hardware and Systems

**Hardware.** We use six different clusters. One is a physical 10-node cluster. Each node has one 2.66GHz Quad Core Xeon processor running 64-bit platform Linux openSuse 11.1 OS, 4x4GB of main memory, 6x750GB SATA hard disks, and three Gigabit network cards. Our physical cluster has the advantage that the amount of runtime variance is limited [30]. Yet, to fully understand the scale-up properties of HAIL, we use three different EC2 clusters, each having 10 nodes. For each of these three clusters, we use different node types (see Section 6.3.3). Finally, to understand how well HAIL scales-out, we also consider two more EC2 clusters: one with 50 nodes and one with 100 nodes (see Section 6.3.4). We report the average runtime of three trials for all experiments.

**Systems.** We compared the following systems: (1) Hadoop, (2) Hadoop++ as described in [12], and (3) HAIL as described in this paper. For HAIL, we disable the HAIL splitting policy (`HailSplitting`) in Section 6.4 in order to measure the benefits of using this policy in Section 6.5. All three systems are based on Hadoop 0.20.203 and are compiled and run using Java 7. All systems were configured to use the default block size of 64MB if not mentioned otherwise.

## 6.2 Datasets and Queries

**Datasets.** For our benchmarks we use two different datasets. First, we use the `UserVisits` table as described in [27]. This dataset nicely matches Bob's Use Case. We generated 20GB of UserVisits data per node using the data generator proposed by [27]. Second, we additionally use a `Synthetic` dataset consisting of 19 integer attributes in order to understand the effects of selectivity. It is worth noting that this `Synthetic` dataset is similar to scientific datasets, where all or most of the attributes are integer/float attributes (e.g., the `SDSS` dataset). For this dataset, we generated 13GB per node.

**Queries.** For the `UserVisits` dataset, we consider the following queries as Bob's workload:

Bob-Q1 (selectivity: 3.1 x $10^{-2}$)

```
SELECT sourceIP FROM UserVisits WHERE visitDate
BETWEEN '1999-01-01' AND '2000-01-01';
```

Bob-Q2 (selectivity: 3.2 x $10^{-8}$)

```
SELECT searchWord, duration, adRevenue
FROM UserVisits WHERE sourceIP='172.101.11.46';
```

Bob-Q3 (selectivity: 6 x $10^{-9}$)

```
SELECT searchWord, duration, adRevenue
FROM UserVisits WHERE sourceIP='172.101.11.46' AND
     visitDate='1992-12-22';
```

Bob-Q4 (selectivity: 1.7 x $10^{-2}$)

```
SELECT searchWord, duration, adRevenue
FROM UserVisits
WHERE adRevenue>=1 AND adRevenue<=10;
```

Additionally, we use a variation of query `Bob-Q4` to see how well HAIL performs on queries with low selectivities:

Bob-Q5 (selectivity: 2.04 x $10^{-1}$)

```
SELECT searchWord, duration, adRevenue
FROM UserVisits
WHERE adRevenue>=1 AND adRevenue<=100;
```

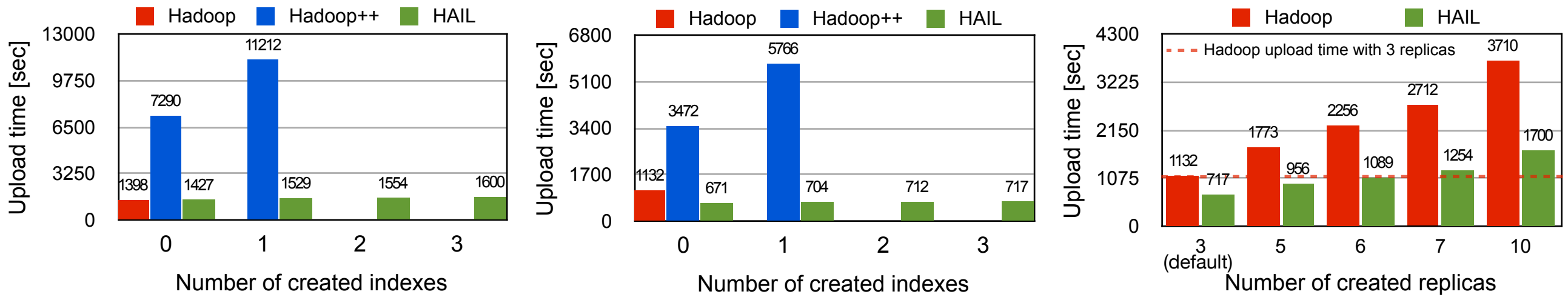

(a) Upload time for `UserVisits` (b) Upload time for `Synthetic` (c) Varying replication for `Synthetic`

**Figure 4: Upload times when varying the number of created indexes (a)&(b) and the number of data block replicas (c)**

**Table 1: `Synthetic` queries.**

| Query | #Projected Attributes | Selectivity |
|---|---|---|
| Syn-Q1a | 19 | 0.10 |
| Syn-Q1b | 9 | 0.10 |
| Syn-Q1c | 1 | 0.10 |
| Syn-Q2a | 19 | 0.01 |
| Syn-Q2b | 9 | 0.01 |
| Syn-Q2c | 1 | 0.01 |

For the `Synthetic` dataset, we use the queries in Table 1.

Notice that for `Synthetic` all queries use the *same* attribute for filtering. Hence for this dataset HAIL cannot benefit from its different indexes: it creates three different indexes, yet only one of them will be used by these queries.

## 6.3 Data Loading

We strongly believe that upload time is a crucial aspect for users while adopting data-intensive systems. This is because most users (such as Bob or scientists) want to start analyzing their data early [18]. In fact, low startup costs are one of the big advantages of standard Hadoop over RDBMSs. Here, we thus exhaustively study the performance of HAIL when uploading datasets.

### *6.3.1 Varying the Number of Indexes*

We first measure the impact in performance when creating indexes. For this, we scale the number of indexes to create when uploading the `UserVisits` and the `Synthetic` datasets. For HAIL, we vary the number of indexes from 0 to 3 and for Hadoop++ from 0 to 1 (this is because Hadoop++ cannot create more than one index). Notice that we only report numbers for 0 indexes for standard Hadoop as it cannot create any indexes.

Figure 4(a) shows the results for the `UserVisits` dataset. We observe that HAIL has a negligible upload overhead of ∼2% over standard Hadoop. Then, when HAIL creates one index per replica the overhead still remains very low (at most ∼14%). On the other hand, we observe that HAIL improves over Hadoop++ by a factor of 5.1 when creating no index and by a factor of 7.3 when creating one index. This is because Hadoop++ has to run two expensive MapReduce jobs for creating one index. For HAIL, we observe that for two and three indexes the upload costs increase only slightly.

Figure 4(b) illustrates the results for the `Synthetic` dataset. We observe that HAIL significantly outperforms Hadoop++ again by a factor of 5.2 when creating no index and by a factor of 8.2 when creating one index. On the other hand, we now observe that HAIL outperforms Hadoop by a factor of 1.6 even when creating three indexes. This is because the `Synthetic` dataset is well suited for binary representation, i.e., in contrast to the `UserVisits` dataset, HAIL can significantly reduce the initial dataset size. This allows HAIL to outperform Hadoop even when creating one, two, or three indexes.

For the remaining upload experiments, we discard Hadoop++ as we clearly saw in this section that it does not upload datasets efficiently. Therefore, we focus on HAIL using Hadoop as baseline.

### *6.3.2 Varying the Replication Factor*

We now analyze how well HAIL performs when increasing the number of replicas. In particular, we aim at finding out how many indexes HAIL can create for a given dataset in the same time standard Hadoop needs to upload the same dataset with the default replication factor of three and creating no indexes. To do this, we upload the `Synthetic` dataset with different replication factors. In this experiment, HAIL creates as many clustered indexes as block replicas. In other words, when HAIL uploads the `Synthetic` dataset with a replication factor of five, it creates five different clustered index for each block.

Figure 4(c) shows the results for this experiment. The dotted line marks the time Hadoop takes to upload with the default replication factor of three. We see that HAIL significantly outperforms Hadoop for any replication factor and up to a factor of 2.5. More interestingly, we observe that HAIL stores six replicas (and hence it creates six different clustered indexes) in a little less than the same time Hadoop uploads the same dataset with only three replicas without creating any index. Still, when increasing the replication factor even further for HAIL, we see that HAIL has only a minor overhead over Hadoop with three replicas only. These results also show that choosing the replication factor mainly depends on the available disk space. Even in this respect, HAIL improves over Hadoop. For example, while Hadoop needs 390GB to upload the `Synthetic` dataset with 3 block replicas, HAIL needs only 420GB to upload the same dataset with 6 block replicas! Thereby, HAIL enables users to stress indexing to the extreme to speed up their query workloads.

### *6.3.3 Cluster Scale-Up*

In this section, we study how different hardware affects HAIL upload times. For this, we create three 10-nodes EC2 clusters: the first uses *large* (*m1.large*) nodes[7], the second *extra large* (*m1.xlarge*) nodes, and the third *cluster quadruple* (*cc1.4xlarge*) nodes. We upload the `UserVisits` and the `Synthetic` datasets on each of these clusters.

We report the results of these experiments in Table 2(a) (for `UserVisits`) and in Table 2(b) (for `Synthetic`), where we display the *System Speedup* of HAIL over Hadoop as well as the *Scale-Up Speedup* for Hadoop and HAIL. Additionally, we show again the results for our local cluster as baseline. As expected, we observe that both Hadoop and HAIL benefit from using better hardware. In addition, we also observe that HAIL always benefits from scaling-up computing nodes. Especially, using a better CPU

[7]For this cluster type, we allocate an additional large node to run the namenode and jobtracker.

**Table 2: Scale-up results**

(a) Upload times for `UserVisits` when scaling-up [sec]

| Cluster Node Type | Hadoop | HAIL | System Speedup |
|---|---|---|---|
| Large | 1844 | 3418 | 0.54 |
| Extra Large | 1296 | 2039 | 0.64 |
| Cluster Quadruple | 1284 | 1742 | 0.74 |
| **Scale-Up Speedup** | 1.4 | 2.0 | |
| Physical | 1398 | 1600 | 0.87 |

(b) Upload times for `Synthetic` when scaling-up [sec]

| Cluster Node Type | Hadoop | HAIL | System Speedup |
|---|---|---|---|
| Large | 1176 | 1023 | 1.15 |
| Extra Large | 788 | 640 | 1.23 |
| Cluster Quadruple | 827 | 600 | 1.38 |
| **Scale-Up Speedup** | 1.4 | 1.7 | |
| Physical | 1132 | 717 | 1.58 |

makes parsing to binary faster. As a result, HAIL decreases (in Table 2(a)) or increases (Table 2(b)) the performance gap with respect to Hadoop when scaling-up (System Speedup). Unlike HAIL, we see that Hadoop does not significantly improve its performance when scaling-up from extra large nodes to cluster quadruple nodes. This is because Hadoop is I/O bound and hence adding better CPUs does not allow it to improve its performance. In contrast, HAIL benefits from additional and/or better CPU cores. Finally, we observe that the system speedup of HAIL over Hadoop is even better when using physical nodes.

### 6.3.4 *Cluster Scale-Out*

At this point, the reader might have already started wondering how well HAIL performs for larger clusters. To answer this question, we allocate one 50-nodes EC2 cluster and one 100-nodes EC2 cluster. We use *cluster quadruple* (*cc1.4xlarge*) nodes for both clusters, because with this node type we experienced the lowest performance variability. In both clusters, we allocated two additional nodes: one to serve as `Namenode` and the other to serve as `JobTracker`. While varying the number of nodes per cluster we keep the amount of data per node constant.

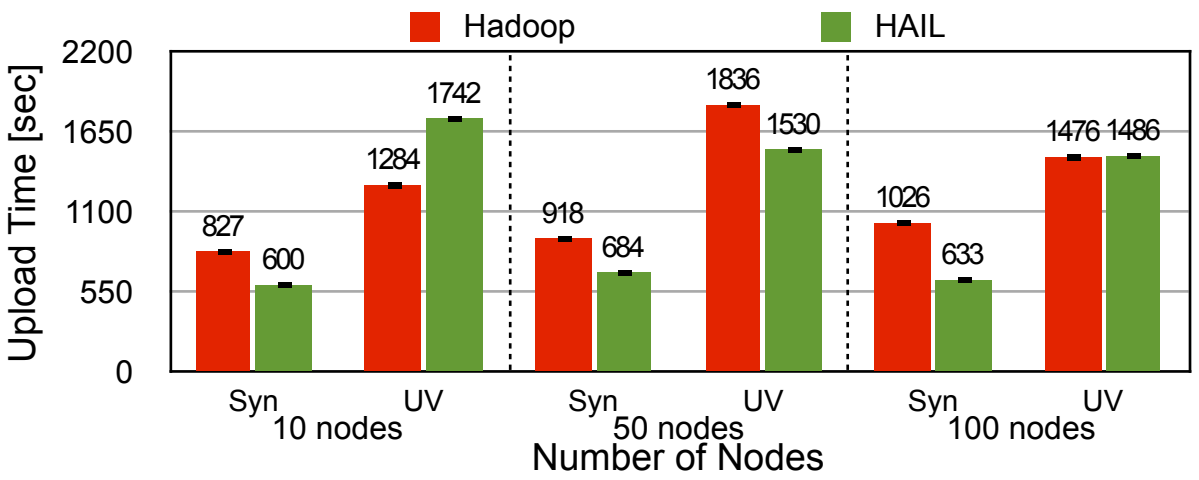

**Figure 5: Scale-out results**

Figure 5 shows these results. We observe that HAIL achieves roughly the same upload times for the `Synthetic` dataset. For the `UserVisits` dataset, we see that HAIL improves its upload times for larger clusters. In particular, for 100 nodes, we see that HAIL matches the Hadoop upload times for the `UserVisits` dataset and outperforms Hadoop by a factor up to $\sim$1.4 for the `Synthetic` dataset. More interesting, we observe that HAIL does not suffer from high performance variability [30]. This is not the case for Hadoop where we observed higher variance. Overall, these results show the efficiency of HAIL when scaling-out.

## 6.4 MapReduce Job Execution

We now analyze the performance of HAIL when running MapReduce jobs. Our main goal for all these experiments is to understand how well HAIL can perform compared to the standard Hadoop MapReduce and Hadoop++ systems. With this in mind, we measure two different execution times. First, we measure the *end-to-end* job runtimes, which is the time a given job takes to run completely. Second, we measure the *record reader* runtimes, which is dominated by the time a given map task spends reading its input data. Recall that for these experiments, we disable the `HailSplitting` policy (presented in Section 4.3) in order to better evaluate the benefits of having several clustered indexes per dataset. We study the benefits of `HailSplitting` in Section 6.5.

### 6.4.1 *Bob's Query Workload*

For these experiments: Hadoop does not create any index; since Hadoop++ can only create a single clustered index, it creates one clustered index on `sourceIP` for all three replicas, as two very selective queries will benefit from this; HAIL creates one clustered index for each replica: one on `visitDate`, one on `sourceIP`, and another one on `adRevenue`.

Figure 6(a) shows the average end-to-end runtimes for Bob's queries. We observe that HAIL outperforms both Hadoop and Hadoop++ in all queries. For `Bob-Q2` and `Bob-Q3`, Hadoop++ has similar results as HAIL since both systems have an index on `sourceIP`. However, HAIL still outperforms Hadoop++. This is because HAIL does not have to read any block header to compute input splits while Hadoop++ does. Consequently, HAIL can start processing the input dataset earlier than Hadoop++ and hence it finishes before.

Figure 6(b) shows the RecordReader times[8]. Once more again, we observe that HAIL outperforms both Hadoop and Hadoop++. HAIL is up to a factor 46 faster than Hadoop and up to a factor 38 faster than Hadoop++. This is because Hadoop++ is only competitive if it happens to hit the right index. As HAIL has additional clustered indexes (one for each replica), the likelihood to hit an index increases. Then, query runtimes for `Bob-Q1`, `Bob-Q4`, and `Bob-Q5` are sharply improved over Hadoop *and* Hadoop++.

Yet, if HAIL allows map tasks to read their input data by more than one order of magnitude faster than Hadoop and Hadoop++, *why do MapReduce jobs not benefit from this?* To understand this we estimate the overhead of the Hadoop MapReduce framework. We do this by considering an ideal execution time, i.e., the time needed to read all the required input data and execute the map functions over such data. We estimate the ideal execution time $T_{\text{ideal}} = \#MapTasks/\#ParallelMapTasks \times \text{Avg}(T_{\text{RecordReader}})$. Here $\#ParallelMapTasks$ is the maximum number of map tasks that can be performed at the same time by all computing nodes. We define the overhead as $T_{\text{overhead}} = T_{\text{end-to-end}} - T_{\text{ideal}}$. We show the results in Figure 6(c). We see that the Hadoop framework overhead is in fact dominating the total job runtime. This has many reasons. A major reason is that Hadoop was not built to execute very short tasks. To schedule a single task, Hadoop spends several seconds even though the actual task just runs in a few ms (as it is the case for HAIL). Therefore, reducing the number of map tasks of a job could greatly decrease the end-to-end job runtime. We tackle this problem in Section 6.5.

### 6.4.2 *Synthetic Query Workload*

Our goal in this section is to study how query selectivities affect the performance of HAIL. Recall that for this experiment HAIL *cannot* benefit from its different indexes: all queries filter on the same attribute. We use this setup to isolate the effects of selectivity.

We present the end-to-end job runtimes in Figure 7(a) and the record reader times in Figure 7(b). We observe in Figure 7(a) that HAIL outperforms both Hadoop and Hadoop++. We see again that even if Hadoop++ has an index on the selected attribute, Hadoop++ runs slower than HAIL. This is because HAIL has a slightly different splitting phase than Hadoop++. Looking at the results in Figure 7(b), the reader might think that HAIL is better than Hadoop++ because of the PAX layout used by HAIL. However, we clearly see in the results for query `Syn-Q1a` that this is not true[9]. We observe that even in this case HAIL is better than Hadoop++. The reason is that the index size in HAIL (2KB) is much smaller than the index size in Hadoop++ (304KB), which allows HAIL to read the index slightly faster. On the other hand, we see that Hadoop++ slightly outperforms HAIL for all three `Syn-Q2` queries. This is because these queries are more selective and then the random I/O cost due to tuple reconstruction starts to dominate the record reader times.

[8]This is the time a map task takes to read and process its input.

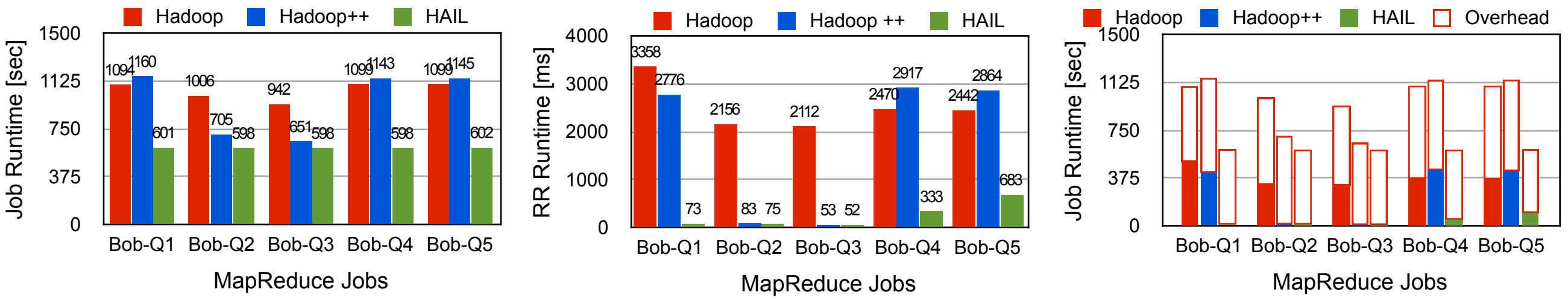

(a) End-to-end job runtimes (b) Average record reader runtimes (c) Hadoop framework overhead

**Figure 6: Job runtimes, record reader times, and Hadoop MapReduce framework overhead for Bob's query workload filtering on multiple attributes**

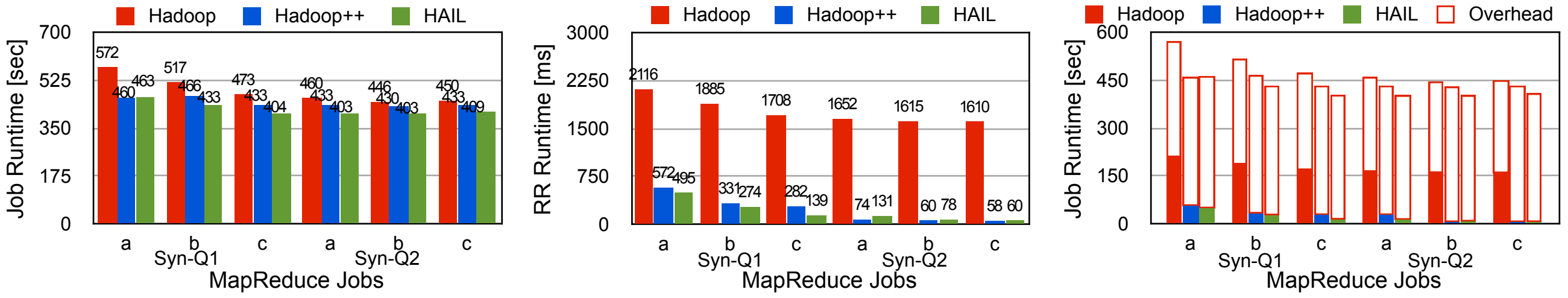

(a) End-to-end job runtimes (b) Average record reader runtimes (c) Hadoop framework overhead

**Figure 7: Job runtimes, record reader times, and Hadoop MapReduce framework overhead for `Synthetic` query workload filtering on a single attribute**

Surprisingly, we observe that query selectivity does not affect end-to-end job runtimes (see Figure 7(a)) even if query selectivity has a clear impact on the RecordReader times (see Figure 7(b)). As explained in Section 6.4.1, this is due to the overhead of the Hadoop MapReduce framework. We clearly see this overhead in Figure 7(c). In Section 6.5, we will investigate this in more detail.

### 6.4.3 *Fault-Tolerance*

In very large-scale clusters (especially on the Cloud), node failures are no more an exception but rather the rule. A big advantage of Hadoop MapReduce is that it can gracefully recover from these failures. Therefore, it is crucial to preserve this key property to reliably run MapReduce jobs with minimal performance impact under failures. In this section we study the effects of node failures in HAIL and compare it with standard Hadoop MapReduce.

We perform these experiments as follows: (i) we set the expiry interval to detect that a TaskTracker or a datanode failed to 30 seconds, (ii) we chose a node randomly and kill all Java processes on that node after 50% of work progress, and (iii) we measure the slowdown as in [12], $slowdown = \frac{(T_f - T_b)}{T_b} \cdot 100$, where $T_b$ is the job runtime without node failures and $T_f$ is the job runtime with a node failure. We use two configurations for HAIL. First, we configure HAIL to create indexes on three different attributes, one for each replica. Second, we use a variant of HAIL, coined HAIL-1Idx, where we create an index on the same attribute for all three replicas. We do so to measure the performance impact of HAIL falling back to full scan for some blocks after the node failure. This happens for any map task reading its input from the killed node. Notice that, in the case of HAIL-1Idx, all map tasks will still perform an index scan as all blocks have the same index.

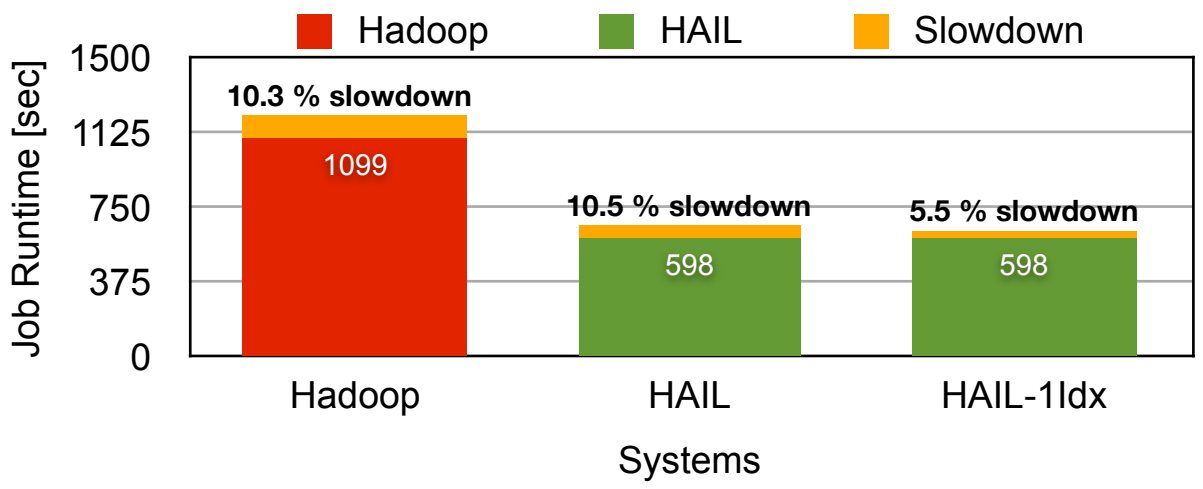

**Figure 8: Fault-tolerance results**

Figure 8 shows the fault-tolerance results for Hadoop and HAIL. Overall, we observe that HAIL preserves the failover property of Hadoop by having almost the same slowdown. However, it is worth noting that HAIL can even improve over Hadoop. This is because HAIL can still perform an index scan when having the same index on all replicas (HAIL-1Idx). We clearly see this when HAIL creates the same index on all replicas (HAIL-1Idx). In this case, HAIL has a lower slowdown since failed map tasks can still perform an index scan even after failure. As a result, HAIL runs almost as fast as when no failure occurs.

## 6.5 Impact of the HAIL Splitting Policy

We observed in Figures 6(c) and 7(c) that the Hadoop MapReduce framework incurs a high overhead in the end-to-end job runtimes. To evaluate the efficiency of HAIL to deal with this problem, we now enable the `HailSplitting` policy (described in Section 4.3) and run again the Bob and Synthetic queries on HAIL.

Figure 9 illustrates these results. We clearly observe that HAIL significantly outperforms both Hadoop and Hadoop++. We see in

[9]Recall that this query projects all attributes, which is indeed more beneficial for Hadoop++ as it uses a row layout.

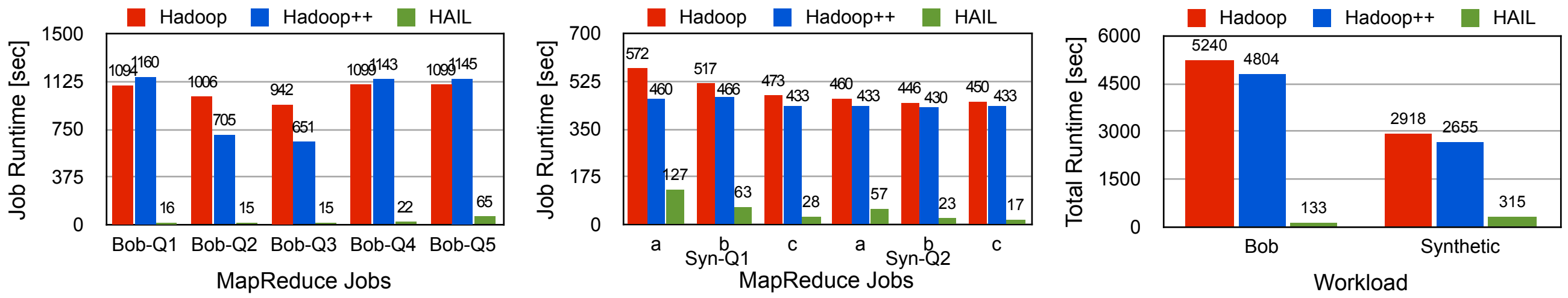

(a) Bob queries (b) `Synthetic` queries (c) Total Workload

**Figure 9: End-to-end job runtimes for Bob and `Synthetic` queries using the `HailSplitting` policy**

Figure 9(a) that HAIL outperforms Hadoop up to a factor of 68 and Hadoop++ up to a factor of 73 for Bob's workload. This is mainly because the `HailSplitting` policy significantly reduces the number of map tasks from 3,200 (which is the number of map tasks for Hadoop and Hadoop++) to only 20. As a result of HAIL Splitting policy, the scheduling overhead does not impact the end-to-end workload runtimes in HAIL (see Section 6.4.1). For the Synthetic workload (Figure 9(b)), we observe that HAIL outperforms Hadoop up to a factor of 26 and Hadoop++ up to a factor of 25. Overall, we observe in Figure 9(c) that using HAIL Bob can run all his five queries 39x faster than Hadoop and 36x faster than Hadoop++. We also observe that HAIL runs all six `Synthetic` queries 9x faster than Hadoop and 8x faster than Hadoop++.

# 7. CONCLUSION

We have presented HAIL (Hadoop Aggressive Indexing Library). HAIL improves the upload pipeline of HDFS to create different clustered indexes on each replica. As a consequence each HDFS block will be available in at least three different sort orders and with different indexes. Like that, in a basic HAIL setup we already get three indexes (almost) for free. In addition, HAIL also works for a larger number of replicas. A major advantage of HAIL is that the long upload and indexing times which had to be invested on previous systems are not required anymore. This was a major drawback of Hadoop++ [12], which created block-level indexes, however required expensive MapReduce jobs to create those indexes in the first place. In addition, Hadoop++ created indexes per *logical* HDFS block whereas HAIL creates different indexes for each *physical* replica. We have experimentally compared HAIL with Hadoop as well as Hadoop++ using different datasets and a number of different clusters. The results demonstrated the high efficiency of HAIL. We showed that HAIL typically creates a win-win situation: users can upload their datasets up to 1.6x faster than Hadoop and run jobs up to 68x faster than Hadoop.

**ACKs.** We would like to thank the anonymous reviewers for their helpful comments. Work supported by M2CI and BMBF.

# 8. REFERENCES

[1] S. Agrawal et al. Database Tuning Advisor for Microsoft SQL Server 2005. *VLDB*, pages 1110–1121, 2004.

[2] A. Ailamaki et al. Weaving Relations for Cache Performance. *VLDB*, pages 169–180, 2001.

[3] S. Blanas et al. A Comparison of Join Algorithms for Log Processing in MapReduce. *SIGMOD*, pages 975–986, 2010.

[4] N. Bruno and S. Chaudhuri. Constrained Physical Design Tuning. *VLDB J.*, 19(1):21–44, 2010.

[5] M. J. Cafarella and C. Ré. Manimal: Relational Optimization for Data-Intensive Programs. *WebDB*, 2010.

[6] S. Chaudhuri et al. Index Selection for Databases: A Hardness Study and a Principled Heuristic Solution. *TKDE*, 16(11):1313–1323, 2004.

[7] G. Chen et al. A Framework for Supporting DBMS-like Indexes in the Cloud. *PVLDB*, 4(11):702–713, 2011.

[8] S. Chen. Cheetah: A High Performance, Custom Data Warehouse on Top of MapReduce. *PVLDB*, 3(1-2):1459–1468, 2010.

[9] D. Dash et al. CoPhy: A Scalable, Portable, and Interactive Index Advisor for Large Workloads. *PVLDB*, 4(6):362–372, 2011.

[10] J. Dean and S. Ghemawat. MapReduce: A Flexible Data Processing Tool. *CACM*, 53(1):72–77, 2010.

[11] J. Dittrich and J.-A. Quiané-Ruiz. Efficient Parallel Data Processing in MapReduce Workflows. *PVLDB*, 5, 2012.

[12] J. Dittrich, J.-A. Quiané-Ruiz, A. Jindal, Y. Kargin, V. Setty, and J. Schad. Hadoop++: Making a Yellow Elephant Run Like a Cheetah (Without It Even Noticing). *PVLDB*, 3(1):518–529, 2010.

[13] M. Y. Eltabakh et al. CoHadoop: Flexible Data Placement and Its Exploitation in Hadoop. *PVLDB*, 4(9):575–585, 2011.

[14] A. Floratou et al. Column-Oriented Storage Techniques for MapReduce. *PVLDB*, 4(7):419–429, 2011.

[15] http://engineering.twitter.com/2010/04/hadoop-at-twitter.html.

[16] Hadoop Users, http://wiki.apache.org/hadoop/PoweredBy.

[17] H. Herodotou and S. Babu. Profiling, What-if Analysis, and Cost-based Optimization of MapReduce Programs. *PVLDB*, 4(11):1111–1122, 2011.

[18] S. Idreos et al. Here are my Data Files. Here are my Queries. Where are my Results? *CIDR*, pages 57–68, 2011.

[19] E. Jahani et al. Automatic Optimization for MapReduce Programs. *PVLDB*, 4(6):385–396, 2011.

[20] D. Jiang et al. The Performance of MapReduce: An In-depth Study. *PVLDB*, 3(1):472–483, 2010.

[21] A. Jindal, J.-A. Quiané-Ruiz, and J. Dittrich. Trojan Data Layouts: Right Shoes for a Running Elephant. *SOCC*, 2011.

[22] W. Lang and J. M. Patel. Energy Management for MapReduce Clusters. *PVLDB*, 3(1):129–139, 2010.

[23] J. Lin et al. Full-Text Indexing for Optimizing Selection Operations in Large-Scale Data Analytics. *MapReduce Workshop*, 2011.

[24] D. Logothetis et al. In-Situ MapReduce for Log Processing. *USENIX*, 2011.

[25] T. Nykiel et al. MRShare: Sharing Across Multiple Queries in MapReduce. *PVLDB*, 3(1):494–505, 2010.

[26] C. Olston. Keynote: Programming and Debugging Large-Scale Data Processing Workflows. *SOCC*, 2011.

[27] A. Pavlo et al. A Comparison of Approaches to Large-Scale Data Analysis. *SIGMOD*, pages 165–178, 2009.

[28] J.-A. Quiané-Ruiz, C. Pinkel, J. Schad, and J. Dittrich. RAFTing MapReduce: Fast recovery on the RAFT. *ICDE*, pages 589–600, 2011.

[29] J. Rao and K. Ross. Making B+-Trees Cache Conscious in Main Memory. *ACM SIGMOD Record*, 29(2):475–486, 2000.

[30] J. Schad, J. Dittrich, and J.-A. Quiané-Ruiz. Runtime Measurements in the Cloud: Observing, Analyzing, and Reducing Variance. *PVLDB*, 3(1):460–471, 2010.

[31] I. Stoica et al. Chord: A Scalable Peer-to-Peer Lookup Service for Internet Applications. *SIGCOMM*, pages 149–160, 2001.

[32] A. Thusoo et al. Data Warehousing and Analytics Infrastructure at Facebook. *SIGMOD*, pages 1013–1020, 2010.

[33] T. White. *Hadoop: The Definitive Guide*. O'Reilly, 2011.

[34] M. Zaharia et al. Delay Scheduling: A Simple Technique for Achieving Locality and Fairness in Cluster Scheduling. *EuroSys*, pages 265–278, 2010.